# Chiral light-matter interactions in hot vapor cladded waveguides

ROY ZEKTZER, ELIRAN TALKER, YEFIM BARASH, NOA MAZURSKI
AND URIEL LEVY

*Department of Applied Physics, The Benin School of Engineering and Computer Science, The Center for Nanoscience and Nanotechnology, The Hebrew University of Jerusalem, Jerusalem, 91904, Israel*
*Corresponding author: ulevy@mail.huji.ac.il*

**Recently, there is growing interest in integrating alkali vapors with nanoscale photonic structures, such as nano-waveguides, resonators and nanoantennas. Nanoscale confinement of electromagnetic fields may introduce a longitudinal electric field component, giving rise to circularly polarized modes which are essential for diverse applications involving vapor and light, such as chirality and non-reciprocity. Hereby, we have designed, fabricated and characterized a miniaturized vapor cell that is integrated with optical waveguides that are designed to generate a peculiar circular-like polarization. Taking advantage of this phenomenon, we demonstrate a spectral shift in the atomic absorption signatures at varying magnetic fields, and significant isolation between forward and backward propagating waves in our atomic-cladded waveguide. Our results pave the way for the utilization of chip-scale integrated atomic devices in applications such as optical isolation and high spatial resolution magnetometry.**

Alkali vapors such as rubidium are being used in various research fields such as quantum information [1–3], nonlinear optics [4–6], magnetometry [7–9] and atomic clocks [10,11]. In the last few years there is a growing interest in miniaturizing Rubidium cells from centimeter scale to micro and nano scale. On top of the obvious advantages of such integration in reducing footprint and cost, many other great qualities are resulted from such an approach. For example, the high confinement allows to observe significant nonlinear effects under very low optical power levels (nano watts) [12–15]. Several miniaturized Rubidium systems have been demonstrated over the last few years, e.g. the atomic cladded waveguide (ACWG) [15,16], hollow core waveguide [17] and coupled atomic-plasmonic systems [14,18]. The combination of strong nonlinear effects and high confinement pave the way for applications such as few photons communication system by all-optical switching [15].

Rubidium is a highly dichroic media due to its strong Zeeman effect and thus has been used to realize a variety of polarization selective and unidirectional devices such as optical isolators [19]. Basically, the Zeeman effect generates circular dichroism, i.e. two orthogonal circular polarizations are experiencing a large difference in their absorption spectrum. This effect has also been used for applications such as frequency stabilization [20], memories [2] and magnetometers [7–9]. While a rectangular waveguide supports quasi linear polarized modes, it has been shown in systems combining waveguides and cold atoms that a strong longitudinal electric field is generated due to the strong field confinement. Thus, one can define the quantization axis by applying magnetic field perpendicular to the propagation direction such that an atom will experience circular polarization interrogation [3]. The absorption lines of such an atom could be controlled by the strength of the magnetic field. This effect has been used for fabricating single photon isolators [21], quantum logic gates [22] and atoms strings as Bragg reflectors [23,24]. This special polarization can also be generated by other photonic devices that confine light to the sub-micron scale such as surface plasmon resonances [25] and nano antennas. It should be mentioned that nano-antennas and meta-materials can also generate left and right circular polarizations and manipulate them by using materials with magneto-optical effect properties [26,27].

Taking advantage of the capability to generate longitudinal polarization component in a nanoscale dielectric waveguide, we have designed and fabricated an integrated nanophotonic-atomic device consisting of a photonic chip integrated with a vapor cell filled with alkali vapor. The ACWG was optimized to generate a circularly polarized light in the evanescent region. Using this chip scale device, we have experimentally demonstrated the interaction of guided mode light with vapor in the presence of magnetic field. In particular, we have measured the transmission spectrum through the ACWG under varying intensities of magnetics field for input-to-output and output-to-input propagation direction and observed a strong circular dichroism. The obtained results are supported by numerical simulations.

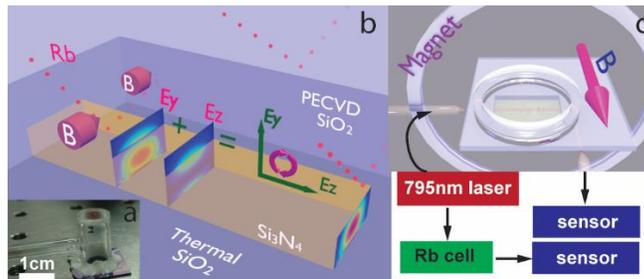

**Fig. 1.** (a) The device. (b) Illustration of the waveguide, the optical modes and magnetic field. The area around the waveguide, where the oxide is etched defines the interaction region between light and vapor. (c) Schematics (not to scale) of the experimental setup

Our hybrid ACWG-magnetic field platform is illustrated in figure 1. The device fabrication process begins with e-beam lithography and reactive ion etching which defines a 3 mm long Si3N4 waveguide, having rib height of 0.25um and width of 0.6 um, sitting on top of thick layer of SiO2. The device is then covered by 2um of SiO2. A ~3mm long interaction region between light and vapor is defined by the top SiO2 using buffer hydrofluoric acid wet etch. A glass chamber is then epoxy bonded to the chip. Next, the device is evacuated, filled with natural rubidium and pinched off. By doing so a chip scale portable device is obtained, as can be seen in figure 1.a. Figure 1.b shows the schematics of the ACWG with polarization components, the TM-like mode profile and magnetic field superimposed. As can be seen, a peculiar polarization is generated owing to the existence of a strong longitudinal component of electric field. Indeed, the polarization plane of the circularly polarized light is parallel to the light's propagation direction. In Fig. 1c we illustrate the experimental setup for the characterization of the fabricated device. We couple light into and out of the chip using lensed fibers and interrogate the D1 line of rubidium using a tunable laser (Toptica ,795nm). We control the magnetic field intensity by positioning a cylinder like magnet at different distances from the device. The magnetic field intensity and orientation is calibrated by using a Hall effect sensor. Temperature is controlled by heating resistors which are attached to the sample. The laser Detuning was calibrated by simultaneously measuring our device alongside with a 7cm long reference cell at room temperature.

First, we have numerically simulated the modes of our ACWG device. The calculated intensity profiles of the longitudinal (z) and the transverse (vertical, y) electric fields modes supported by our waveguide as shown in figure 2 (a, b). As we excite the TM-like mode, the electric field intensity in the x direction is an order of magnitude smaller than the field intensities along the y and z directions. Therefore, the contribution of this field component is neglected. It can be seen that in the region of the evanescent wave above the waveguide, which is the major region contributing to light-vapor interactions, the fields intensities of the two field components are similar. This combined with the $\pi/2$ phase shift between the two fields results in an unusual circularly like polarized light. We have confirmed the existence of circular polarization by a finite-difference-time-domain (FDTD) simulation and plotted the field orientation 30 nm above the waveguide (figure 2c). We can decompose the field as $\Psi = a \cdot |E_y\rangle + b|E_z\rangle = (a-b)/\sqrt{2}|R\rangle + (a+b)/\sqrt{2}|L\rangle$. a and b are the amplitudes of the transverse ($|E_y\rangle$) and the longitudinal ($|E_z\rangle$) electric field respectively, $|R\rangle$ and $|L\rangle$ are the right and left circular based electric fields. The ratio between the magnitude of the electric field in the circular polarizations $(a-b)/(a+b)$ defines the purity of circular polarization where 0 corresponds to purely circularly polarized light, 1 is transverse linear polarization and -1 is longitudinal linear polarization. In figure 2d we plot the ratio across the mode, with the polarization ellipse superimposed.

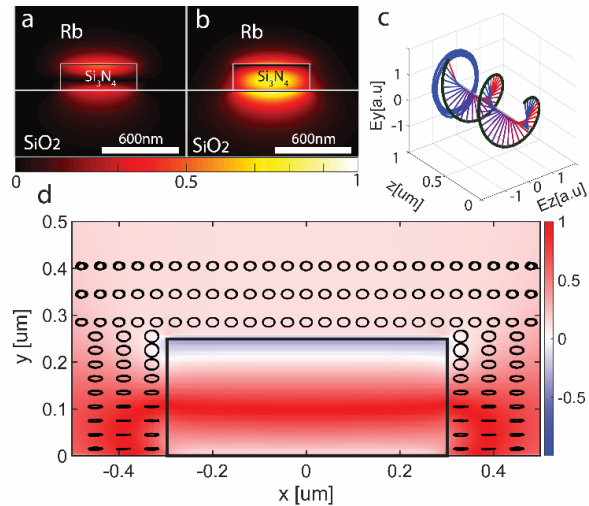

**Fig. 2**. (a,b) Magnitude of the calculated longitudinal and the transverse (vertical) electric fields. The waveguide structure is superimposed. Bright corresponds to high intensity. (c) FDTD simulation showing the evolution of polarization along the propagation axis. We have chosen a transverse coordinate located 30 nm above the top of the waveguide in the vertical direction and centered in the horizontal direction. The circular nature of polarization is evident. (d) In color- the ratio between left handed polarization and right-handed polarization (0 corresponds to circularly polarized light). Outside of the waveguide, in the region of interaction between light and atoms, we have superimposed the polarization ellipse (as in block c). The polarization within the waveguide is not shown, as no light-vapor interactions take place within the waveguide core.

One can clearly see that most of the region where the mode interacts with the atom the polarization is circularly polarized. In fact, by decomposing the optical mode into circular polarization basis, we have numerically found that about 93% of the optical power which resides in the evanescent tail of the optical mode is projected into one of the circular eigenstates, i.e. we have circular polarization purity of 93% in the evanescent tail of the optical mode.
Next, we simulate the effect of the application of external DC magnetic field on the D1 transitions by solving the eigenvalues of the Hamiltonian of the atom under perturbation of magnetic field [28]. We found the energy shift of each Zeeman sub-level and added a Voigt profile to each transition to account for Doppler and transit time broadening [29]. In previous works [29] we have shown that for the case of ACWG a larger Doppler broadening is obtained as compared to the free space due the effective refractive index of the mode being larger than unity ($n_{eff} = 1.56$) which results in Doppler broadening of about 780MHz. Furthermore, the finite interaction time of the atom with the evanescent field results in a transit time broadening of about 300MHz. Combining both broadening processes results in a signal with approximately 1.1 GHz linewidth. We have added a more detailed explanation on the broadening and on finding the eigenvalues in the supplementary. In figures 3a-3d, we present the simulated (top panel) and measured (bottom panel) transmission spectra of our ACWG device with several values of applied magnetic field (colored lines) for left and right-handed polarizations.

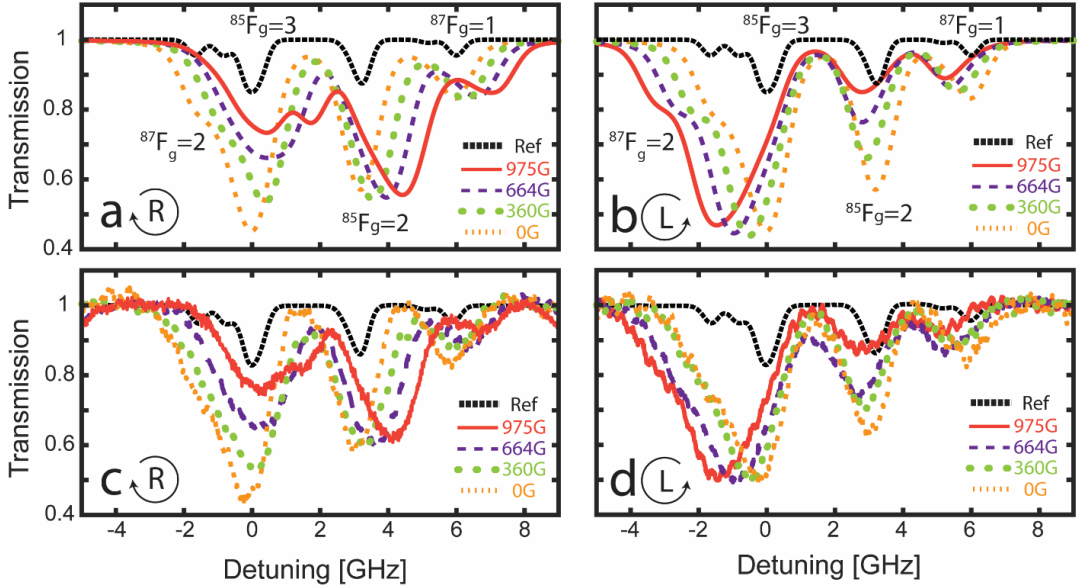

**Fig. 3**. (a,b) Simulated transmission spectra of the ACWG with different magnetic fields, a- right handed, b-left handed polarization. (c,d) Measured transmission spectra of the ACWG with different magnetic fields, c- right handed, d-left handed polarization. Spectroscopy data was captured at temperature of 110°. Spectroscopy data from the reference cell is captured at temperature of 25° and presented not in scale.

The polarization can be switched either by flipping the magnet or by switching between the input and the output coupling to the device. For the data presented in figure 3 we have used the latter, in support of the optical isolation application. We have excited the TM-like mode supported by our waveguide using polarizing maintaining lensed fibers to ensure the launching of circularly polarized mode. The coupling loss from the external fiber to our ACWG was measured to be ~15dB from which we deduce that the power in the waveguide is ~5nW, an order of magnitude less than the saturation power [29]. The device was heated to 110° by using resistors, due to the small size of the device we could heat it without heating the magnet beyond its operation temperature (80°). As a reference, we have recorded simultaneously the Rubidium absorption line in a reference Rubidium cell (black lines). From the result one can notice four signatures of absorption due to the combined transitions from the two ground levels of Rb85 and the two ground levels of Rb87. Furthermore, one can clearly observe a significant change in the transmission spectrum as a function of the applied magnetic field. There is a distinct shift in absorption signatures. The direction of the shift is determined by the polarization. In figure 3, panels a, c ($\sigma_+$) one can clearly observe a blue shift of the absorption spectrum. This shift is a direct result of the Zeeman effect. For example, at 975G there is a ~ 1.2GHz blue shift of the absorption signature which is associated with the F=2 ground level of Rb85. On the other hand, in panels b and d ($\sigma_-$) we can clearly observe a red shift of the absorption spectrum. For example, at 975G there is a ~ 1.5GHz red shift of the absorption signature which is associated with the F=3 ground level of Rb85. Furthermore, we can observe both in the simulation (a) and the measurement (c) a red shift of the absorption signatures resulted from the F=2 ground level of Rb87 and the F=3 ground level of Rb85. These signatures cannot be fully resolved due to Doppler and transit time broadening, thus resulting a saddle pattern in their spectral response. Generally speaking, there is a good agreement between our simulation and measurement due to high purity of circular light in the evanescent interaction region. We also observe what seems to be a contrast reduction of the other (F=2 for $\sigma_-$ and F=3 for $\sigma_+$) absorption signature of Rubidium 85. This is due to the larger difference in Zeeman shift of the sublevels. While there is a good agreement between our simulations and measurements, there are small differences resulted mostly from facet reflections, introducing etalon effects in our devices.

In figure 4 we present the measured transmission of light through the device in right and left polarized configurations, for three values of magnetic field. To estimate the capability of the device to act as an isolator, we define an isolation ratio given as the ratio of absorption between the two orthogonal circular polarizations. As can be seen, a magnetic field of 359 Gauss (panel a) do not yield much isolation. By increasing the magnetic field strength to 664 Gauss, one obtains a higher isolation ratio of about 2.5. Finally, by increasing the magnetic field to 975 Gauss, a very high isolation ratio of ~10 is observed. It should be noted that in addition to isolation ratio, another important parameter is the absolute value of isolation, where isolation in the level of few tens of dB is typically needed. While our current device shows absorption of about 50 percent (3 dB), we have already shown in previous works that contrast can be greatly improved by operating at higher temperatures [29] resulting in higher atom density and by increasing the length of the active area [15]. The current operating temperature of the device is limited due to the limited temperature operation of the glue used for bonding and of the magnets. In the future we intend to further improve the extinction ratio by the use of anodic bonding and by using chip scale magnets. This will allow to operate at even higher temperatures. Furthermore, longer waveguide will be used. These changes should allow further improvement in the absolute value of isolation.

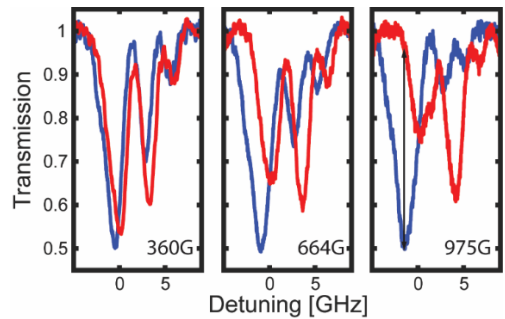

**Fig. 4**. Measured transmission spectra of the ACWG with different magnetic fields. Each panel compares between the left (blue) and the right-handed polarization (red). We observe optimal isolation at 975G.

With such modifications, our isolator can be implemented within communication system operating in the few photon regime. The enhanced broadening of the rubidium transitions in our system play a positive role as it supports higher operation bandwidth. Another possible application for our device is in the field of magnetometry. As can be seen in figure 3 the device is indeed sensitive to magnetic fields. While the precision of the magnetic field sensing is expected to be somewhat limited due to linewidth broadening, the active region of detection in our device is confined to about 100 nm (the decay length of the evanescent wave) offering a tremendously high spatial resolution magnetic sensor. The high resolution makes this system a possible candidate for imaging magnetometry. While we are not yet able to achieve nano scale resolution that can be achieved with a single NV center [30] or magnetic field resolution achieved in large scale vapor cells [8] we believe that our device provides an important step [7] toward nano scale high resolution magnetometry. As such, the presented approach has the potential to serve as a bridge between the above mentioned approaches, providing both high spatial resolution and high magnetic field resolution using chip scale technology. Future designs may also include metallic wires integrated on the chip, such that alternating magnetic fields can be applied.

In conclusion, we designed, fabricated and experimentally characterized a chip scale non-reciprocal ACWG device based on the integration of dielectric waveguide with alkali vapors and demonstrated the control over its transmission by the application of magnetic fields. The co-existence of a strong longitudinal field component which is in quadrature to the transverse electrical field result in a peculiar circularly polarized mode with the polarization plane parallel to the propagation direction. By the application of magnetic field perpendicular to the polarization plane, an effective Faraday configuration is achieved, giving rise to $\sigma^+$ and $\sigma^-$ transitions, depending on the polarization and the propagation direction. The experimental results, which are supported by numerical simulations, indicate isolation between the forward and the backward propagating beams, with a bandwidth of about 1 GHz. For a magnetic field of 975 Gauss, very high isolation ratio is observed. Higher values of magnetic fields can be applied in the future by optimizing the geometry of the chip and the magnet. The presented platform can be a base for various applications such as isolators, high spatial resolution magnetometers and magneto optic modulators. This platform can also be used for the integration of many rubidium-magnetic based applications such as optical memories and frequency stabilization.

**Funding.** The research was supported by the European Research Council (ERC-LIVIN 648575), and by the Israeli ministry of science and technology.

See Supplement 1 for supporting content.

# Chiral light matter interaction in hot vapor cladded waveguides: supplementary material

ROY ZEKTZER, ELIRAN TALKER, YEFIM BARASH, NOA MAZURSKI AND URIEL LEVY

*Department of Applied Physics, The Benin School of Engineering and Computer Science, The Center for Nanoscience and Nanotechnology, The Hebrew University of Jerusalem, Jerusalem, 91904, Israel*
*Corresponding author: ulevy@mail.huji.ac.il*

### A. Calculation of Rubidium transition susceptibility in the evanescent region

In order to evaluate the effect of magnetic field on our absorption first we need to evaluate the effect broadening mechanisms will have on our line shape. The interaction of light with the Rubidium in our system is modelled in the following way, adopting the formalism which is used to calculate the susceptibility of a quantum emitter that is reflected from a surface with decaying field [1,2]. Essentially, the susceptibility of Rubidium is estimated according to:

$$\chi(\nu) \sim \int_{-\infty}^{\infty} dv_z \int_{0}^{\infty} dv_T \frac{W_D(v_z, v_T)}{-2\pi(v_0 - \nu) - k_z v_z - i(\gamma - i k_T v_T)} \quad (s1)$$

Where, $v_0$ is the frequency of a single transition, $v_z$ and $v_T$ are the velocities of the Atom along the propagation and transverse direction, respectively, $W_D$ is the Boltzmann thermal velocity distribution, $k_z$ and $k_T$ are the wave numbers of the mode in the propagation direction and at the transverse direction respectively. They relate to the wavenumber via $k_z^2 + k_T^2 = k^2$. $\gamma$ is the absorption line natural linewidth. From the susceptibility we can deduce the refractive index of Rubidium according to $n = \sqrt{1+\chi}$. In order to do so, we first need to deduce the mode wave vector. We have used Lumerical mode solver to find the effective index of the mode ($n_{eff}$) to be 1.56. The wave number of the mode in the propagation direction is $k_x = 2\pi/\lambda \cdot n_{eff}$ and in the transverse direction is ($\sqrt{(2\pi/\lambda)^2 - k_x^2}$). Once we have evaluated the mode wave number we can solve eq 1 for each transition and receive the rubidium imaginary and real parts of the refractive index as can be seen in figure s1a,b respectively. About 10% of the optical mode interacts with the Rb vapor in our device. As a result, the absorption of the optical mode is about 10% of the absorption of the Rb transition. For example, if the Rb cladding has an imaginary index of 2E-4, the mode will have an effective imaginary index of 2E-5. The transmission of our device is described by $T = \exp(-\alpha \cdot L \cdot IF)$ where $\alpha$ is the absorption coefficient deduced from the imaginary index of rubidium, L is the length of the device (3 mm) and IF is the interaction factor of the TM mode we excite with the Rb vapor (~10%). In figure s1.c we plot the simulated transmission.

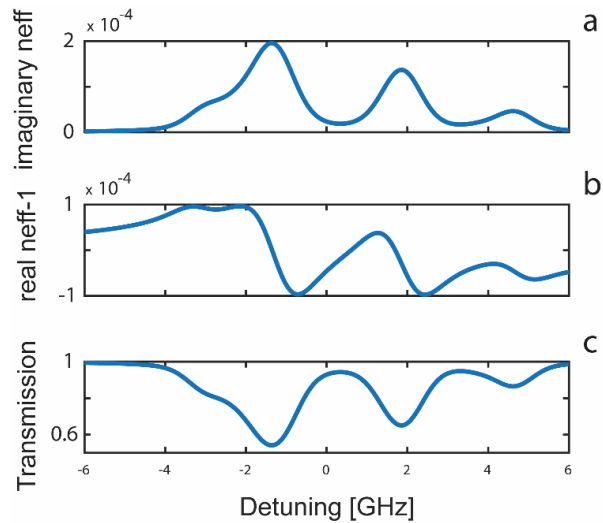

**Fig**. s1. (a-b) Simulated imaginary and real part of the refractive index of Rb[85]. (c) simulated transmission of our ACWG

### B. Finding the eigenvalues of the atom Hamiltonian under the influence of magnetic field

The Hamiltonian of the atom with magnetic field B is described by

$$H = H_0 + \left(-\frac{\mu_B}{\hbar}\right) \cdot B \cdot (L + g_s S + g_I I) \quad (s2)$$

Where S is the spin angular momentum and L is the orbital angular momentum, I is the nuclear spin, g is the Lande g sactors, B is the applied magnetic field, $\mu_B$ is the Bohr magneton and $H_0$ is the un perturbated Hamiltonian. In order to understand the behavior of the atoms with the presence of magnetic field we need to solve the eigenvalues of the Hamiltonian. By solving the eigenvalues of the Hamiltonian, we found the energy shift of each Zeeman sub-level. The direction of the magnetic field defines the quantization axis. Therefore, we can us the unperturbed atomic states vector $|F, m_F\rangle$, under this representation the expectation value of the diagonal elements in the Hamiltonian is $\langle F, m_F | H | F, m_F \rangle = E_0(F) - \mu_B g_F m_F B_Z$. Where $E_0$ is the energy sublevels of the $|F, m_F\rangle$ and $g_F$ is the associated Lande factor. The expectation value of the off-diagonal matrix elements may be non-zero only between $\Delta F = \pm 1$ and $\Delta m_F = 0$ sub levels. These values are evaluated by the following expression:

$$\langle F-1, m_F | H | F, m_F \rangle = \langle F, m_F | H | F-1, m_F \rangle =$$

$$\frac{-\frac{\mu_B}{2}(g_J - g_I) B_z \left([(J+I+1)^2 - F^2] \cdot [F^2 - (J-I)^2]\right)^{0.5} (F)^{-0.5}}{(F^2 - m_F^2)^{-0.5} (F(2F+1)(2F-1))^{0.5}}$$

Now we consider applying laser light, the laser lights influence on the atom is taken into account by interaction term coupling electric dipole and the oscillating laser electric dipole $-d \cdot E$. Therefore, the $|g\rangle -> |e\rangle$ transition dipole moment for an atom interacting with longitudinal magnetic field is proportional to

$$|\langle e | D_q | g \rangle| \propto \sum_{F_e, F_g} c_{F_e' F_e} a(F_e, mF_e; F_g, mF_g; q) c_{F_g' F_g}$$

With

$$a(F_e, mF_e; F_g, mF_g; q) =$$

$$\frac{(-1)^{1+I+J_e+F_e+F_g-mF_e}}{(2J_e+1)^{-0.5}(2F_e+1)^{-0.5}(2F_g+1)^{-0.5}} \cdot \begin{pmatrix} F_e & 1 & F_g \\ -m_{F_e} & q & m_{F_g} \end{pmatrix} \cdot \begin{Bmatrix} F_e & 1 & F_g \\ J_q & I & J_e \end{Bmatrix}$$

Where the parentheses and the curly brackets denote the 3-j and 6-j coefficients, respectively. q is the polarization of the light (e.g for right handed circular polarization light ($\sigma_+$) q=1)

___

### C. Simulating the ACWG response taking into account the transitions shifts and broadening effects.

In figure s2 we present the simulation results adding the rubidium 85 (green) and 87 (blue) transitions. These simulations are calculated for temperature of 70C.

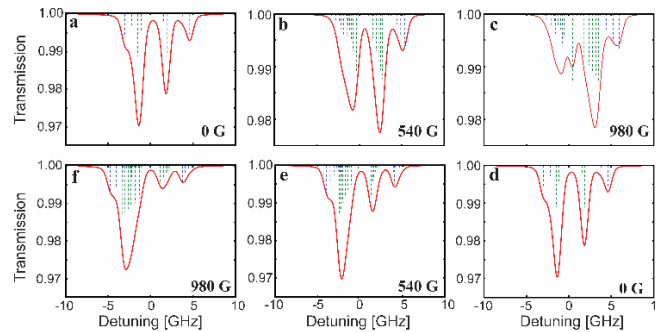

**Fig**. s2 Simulated transmission spectra of the ACWG with different magnetic fields (Red dotted lines), a-c right handed, d-f left handed polarization. The additional green lines represent the Rb85 transitions and the additional blue lines represent the Rb87 transitions.